\documentclass[10pt,twocolumn,aps,prb,showpacs]{revtex4-1}
\usepackage[T1]{fontenc}
\usepackage{verbatim}
\usepackage{graphicx}
\usepackage{graphics}
\usepackage{upgreek}
\usepackage{array}
\usepackage{mathrsfs}
\usepackage{amsmath}
\usepackage{pstricks}
\usepackage{dsfont}
\usepackage{amssymb}
\usepackage{lipsum}
\usepackage{color} 
\newcommand\ii{\mathbf{i}}
\newcommand\cor[1]{\langle #1 \rangle}

\newcommand\rr{\textbf{r}}
\newcommand\qq{\textbf{q}}

\newcommand\MZ{\mathscr{Z}}

\newcommand\GP{\mathcal{G}}
\newcommand\CP{\mathcal{C}}

\newcommand\HP{\mathcal{H}}
\newcommand\ket[1]{| #1 \rangle}

\newcommand\MBC{\boldsymbol{\mathcal{C}}}
\newcommand\Bphi{\boldsymbol{\Phi}}

\newcommand\BJ{\boldsymbol{J}}

\newcommand\iinfty{\displaystyle\int_{-\infty}^{\infty}\,}
\newcommand\el{\textrm{el}}
\newcommand\env{\textrm{env}}
\newcommand\res{\textrm{osc}}
\newcommand\eff{\textrm{eff,e}}
\newcommand\bos{\textrm{eff,b}}

\newcommand\sgn{\text{sgn}}

\begin{document}			
\title{Dynamical Coulomb Blockade in an interacting 1D system coupled to an arbitrary environment}
\author{J.-R. Souquet}
\author{I. Safi}
\author{P. Simon}
\affiliation{Laboratoire de Physique des Solides, Universit\'e Paris-Sud, 91405 Orsay, France}
 \date{\today}

\begin{abstract}
We study the out-of-equilibrium transport in a Tomonaga-Luttinger liquid containing a weak or a tunneling barrier coupled to an arbitrary electromagnetic environment. This applies as well to a coherent one-channel non-interacting conductor with  a transmission coefficient close to one or to zero. We derive formal expressions for the current and finite-frequency (FF) noise at arbitrary voltages, temperatures and frequency-dependent impedance $Z(\omega)$ in the regimes of weak  and  strong backscattering. We show that these two regimes are no longer related by duality at finite frequency. We then carry explicit computations of the nonlinear conductance and FF noise when $Z(\omega)$ describes an harmonic oscillator such as a LC circuit or a cavity.
\end{abstract}

\maketitle
\section{Introduction}

In the last two decades, a lot of attention has been brought both theoretically and experimentally to the back action of an electromagnetic environment on the electronic transport properties of a coherent conductor.
The majority of these studies was devoted to the case where the conductor reduces to a tunnel junction.\cite{panyukov_zaikin,odintsov,nazarov_89,IN,girvin_90,Devoret,Hofheinz} The tunneling of an electron provokes sudden voltage variations at the edge of the environment which excite its electromagnetic modes. These, in turn, affect the charge transfer and reduce the conductance of the coherent conductor in a non-linear fashion. Such a phenomenon, called the dynamical Coulomb blockade (DCB), has been best studied within the well established P(E) theory \cite{panyukov_zaikin,odintsov,nazarov_89,girvin_90,Devoret,IN} which accounted for experimental results obtained both in normal \cite{delsing89,cleland90} and superconducting \cite{holst94,Hofheinz} tunnel elements connected to an electromagnetic environment. \\
More recently, an increasing interest has been devoted to a well transmitting conductor, with transmission coefficient $\tau$ close to unity, where one expects naively the DCB to become negligible due to important charge fluctuations. Indeed,  DCB vanishes for a perfect transmission, $\tau=1$. This is also the case for the bare shot noise of the conductor. Interestingly, the DCB reduction of the conductance was shown to be proportional to shot noise, establishing its link to charge granularity.\cite{golubev_zaikin_01,yeyati,pierre} Nevertheless, such a relation has been derived perturbatively with respect to a small environmental impedance, and shows a divergence at low energies. Further theoretical progress became possible when the impedance reduces to a pure resistance $R$ (at low frequencies below the RC frequency), leading to a counterintuitive result: even a small resistance operates a strong back action on a well-transmitting conductor at low energies, leading to the suppression of the conductance at zero voltage.\cite{Kindermann,Safi} Furthermore Ref. [\onlinecite{Safi}] has allowed to fill the gap for arbitrary values of both the environmental resistance $R$ and the transmission $\tau$ by mapping this problem to the one of an impurity in a Tomonaga-Luttinger liquid (TLL) with an interaction parameter
\begin{equation}K=\frac 1{1+R/R_q},\label{key}
\end{equation} where $R_q=h/e^2$ is the quantum of resistance. In other words, coupling the conductor to an ohmic environment is locally equivalent to introducing interactions between electrons and its effect on transport cannot be treated as a perturbation. Such an equivalence has opened the path to emulate the TLL physics and tune the interaction parameter $K$ through $R$. This has been recently backed up by state of art experiments which enabled to engineer the right ohmic environment, vary the resistance in situ, and study its strong back-action on a quantum point contact (QPC), leading to both successful realizations of an impurity in a  TLL as well as further understanding of the DCB phenomenon.\cite{pierre,jezouin} This novel path has also led, through tuning $R$, to control the quantum phase transition which is predicted to occur when interacting leads are coupled to a double-barrier resonant level structure.\cite{florens07,mabrahtu,florens12} 

By exploiting the mapping to an impurity in a TLL, the link between DCB and noise was made rigorous by deriving a  relation between the nonlinear conductance and the out-of-equilibrium noise in the presence of the environment, which is non-perturbative both at arbitrary transmission $\tau$ and resistance $R$.\cite{Safi} Beyond zero-frequency noise, finite frequency (FF) noise \cite{BlanterButtiker} offers valuable information on the dynamics of excitations and can probe even better the underlying model. It now has become  accessible experimentally.\cite{PortierFF,Basset1,Basset2} In particular, as interactions have striking effects in one dimension, FF noise turns out to be  strongly affected as well. Initially  the symmetrized FF noise was studied for fractional charge tunneling in a constriction between two edge states in the fractional quantum effect (FQHE).\cite{chamon} The underlying model is a TLL with a weak backscattering (WBS) center. At low energy, one has to deal with the limit of a pinched off  constriction, thus of strong backsckattering (SBS), where only electrons tunnel, for which the symmetrized FF noise thus computed was claimed to be dual. Nevertheless, these results were revisted in  [\onlinecite{tztl}], showing that the FF symmetrized noise contains additional contributions in the WBS regime. This has required to develop a formalism based on Keldysh techniques and suited to treat an impurity in a TLL, as well as to one-dimensional systems with Coulomb interactions having an arbitrary range. More relevant to current experiments is the FF non-symmetrized noise, which gives access to both the emission and absorption spectrum. Its first study  in a TLL with an impurity was performed in Ref. [\onlinecite{BenaSafi}], and applies both to the FQHE and a conductor connected  to a resistance $R$ after Eq.(\ref{key}); it was shown explicilty that duality between the WBS and SBS is broken. Other studies have followed, such as in non chiral TLLs,\cite{Bena,recher_FF_noise,souquet}  and more recently in a one channel conductor connected to a resistance $R=h/e^2$, which corresponds to a TLL with an interaction parameter $K=1/2$.\cite{zamoum_12} 

Another important issue one needs to explore is the interplay between intrinsic Coulomb interactions inside  the conductor and the electromagnetic environment in series. In a one-channel conductor, interactions lead to collective excitations, and the picture of an electron tunneling from an occupied state to an available state on the other side of the barrier is no longer valid. Indeed, the problem of an impurity in a TLL with an interaction parameter $K$ in series with a resistance $R$ can be mapped at low energy  to an impurity in a TLL with parameter $K'= K/(1+KR/R_q)$,\cite{Safi}  which generalizes Eq.(\ref{key}). 
A promising extension of this mapping has been recently made in Ref.[\onlinecite{jezouin}] in order to go beyond the restriction to an ohmic environment: a conductor with an arbitrary transmission in series with an arbitrary frequency-dependent  impedance $Z(\omega)$ can be mapped, formally, to a one-dimensional wire with finite-range Coulomb interactions. 

The aim of this paper is to describe the non-equilibrium transport properties of a weak or strong scatterer embedded in an interacting conductor, described by a TLL, in series with an arbitrary impedance $Z(\omega)$. We consider specifically perturbative regimes in the SBS and WBS, for $\tau$ close to one. The SBS is useful both to explore the low energy physics when one starts from $\tau$ close to one, but also in the tunneling regime when $\tau\ll1$.  We show that in both the SBS and the WBS limits, the DC current obeys a $P(E)-$ like formula, 

\begin{equation}
I(V)=e\lambda^2\iinfty dE P(E)\left(\Gamma_K(E+eV)-\Gamma_K(E-eV)\right),
\end{equation}
as well as the differential conductance:
\begin{equation}\label{eq:fdtgen}
G(V)=\frac{dI(V)}{dV}=
\iinfty d\omega P(\omega)\partial_\omega S_i(-\omega).
\end{equation}
In these expressions, $\Gamma_K$ can be interpreted as the tunneling or the backscattering probability at the impurity, $S_i(eV,\omega)$ is the finite frequency noise at frequency $\omega$ emitted by the QPC {\em without} an environment, $P(\omega)$ the probability for the environment to absorb a quanta of energy $h\omega$, and $\lambda$ a small parameter that describes its coupling to the TLL.
This provides a relation between the DCB and the bare noise of the TLL which is perturbative with respect to $1-\tau$ or $\tau$, but not with respect to the impedance $Z(\omega)$, neither the interactions in the TLL, as $K$ is arbitrary. Thus it extends the form derived in Ref. [\onlinecite{Gab_arxiv}] (see also [\onlinecite{Gab08}])  for a noninteracting conductor (therefore  at $K=1$) and  for a tunneling barrier which corresponds to $\tau\ll 1$.
Then, we derive   the  perturbative formal expressions of the FF noise when the environment corresponds to a LC oscillator. We then show explicitly that duality between WBS and SBS is violated.

By letting $K=1$, our present study also allows to treat a non-interacting conductor, such as a QPC in a two-dimensional gas, connected to an LC resonator. In this case, we derive novel results for the FF noise and the feedback effects when $\tau$ is close to one.
For $K<1$, the TLL could simulate a noninteracting conductor in series with a resistance $R$, related to $K$ via Eq.(\ref{key}).  It could as well simulate edge states in the FQHE where a QPC is created by applying gate voltage. For a realistic interacting quantum wire of length $L$,  one can expect some of the features to remain qualitatively valid  provided all frequency scales are above the typical frequency $u/L$  where $u$ is the plasmon velocity. The frequency $u/L$ is of the order of $10 GHz$, which is currently accessible experimentally.\cite{PortierFF,Basset1,Basset2} For simplicity, we will still use QPC to refer commonly to any of these systems, in absence or presence of interactions, modeled by a TLL with an impurity.

This paper is organised as follows: In Sec. II we recall a few results from the $P(E)$ theory that will be extended to the case of interacting electrons in Sec. III. In particular, we will derive the current and the FF noise as a function of temperature and bias in both transmitting regimes. In Sec. IV, the formal expressions thus obtained will be explicited in the case of an harmonic oscillator.

\section{P(E) theory for a non-interacting wire}

\begin{figure}
\vspace{-1.5cm}
\begin{pspicture}(0,0)(7,6)
\rput[bl](-.5,0.3){\includegraphics[width=0.45\textwidth]{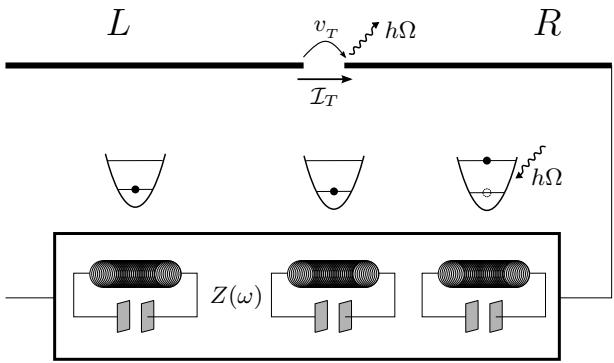}}
\rput[b](1.,4.65){\Large $L$}
\rput[b](6.7,4.65){\Large $R$}
\rput[b](4.75,4.6){$h\Omega$}
\rput[b](6.7,2.65){$h\Omega$}
\rput[b](3.75,3.7){$\mathcal I_T$}
\rput[b](3.78,4.61){$v_{_T}$}
\rput[b](2.57,1.){$Z(\omega)$}
\end{pspicture}
\caption{\label{schemas} Sketch of a QPC acting as a tunnel barrier on a one channel quantum conductor coupled to an environment. As an electron tunnels through the QPC, part of its energy $h\Omega$  is absorbed by the environment characterized by an impedance $Z(\omega)$.}
\end{figure}

In this section, we first recall a few key results of the P(E) theory\cite{IN} which we then apply to the case of an harmonic oscillator.

One of the main success of the P(E) theory is to give a clear picture in terms of probability of the tunneling rate in presence of an electromagnetic environment. The current is expressed as the probability for an electron of energy $eV$ to tunnel through the barrier and to exchange a photon of energy $E$ with the environment (See Fig. \ref{schemas} for a shematic illustration of the P(E) theory applied to a QPC). This probability, denoted as $\Gamma_{\rm env}(V)$ below, is given by a convolution between the P(E) function and $\Gamma_{(K=1)}=\Gamma_1$, the tunneling probability to the right (in absence of the environment) and reads:

\begin{equation}
\Gamma_{\rm env}(V)=\frac{1}{e^2R_T}\left[\Gamma_1*P\right](eV),
\label{eq:PET}
\end{equation}
where $R_T$ is the tunneling resistance, $[\cdot*\cdot]$ denotes the convolution product, $\Gamma_1(eV)=[f*(1-f)](eV)$ is the probability for an electron to tunnel from an occupied state to an unoccupied state on the other side of the junction, and $f(E)$ is the Fermi distribution. (In this paper we set $\hbar=1$, except for the energy of a photon of frequency $\Omega$ for which we keep the notation $h\Omega$.) Another important result is the analytical expression of P(E) for a harmonic environment:
\begin{eqnarray}
&P(E)&=\frac{1}{2\pi}\displaystyle \int dt \exp\left(J(t)-iEt\right),~~~{\rm with}\\
&J(t)&=\displaystyle \int~
d\Omega\Re \left[Z(\Omega)\right]\left(\CP_{\res}(t)-\CP_{\res}(0)\right).
\label{eq:Jtfunction}
\end{eqnarray}
$\CP_{\res}(t)=\cor{\varphi (t)\varphi(0)}$ is the "lesser" Green function of the harmonic oscillator of frequency $\Omega$. If the oscillator corresponds to a LC resonator, $\varphi$ refers to the phase, such that the voltage drop across the LC element is given by $U_{env}(t)=\partial_t\varphi(t)/e$.  Thus $\CP_{\res}(t)$ can be deduced from the equilibrium fluctuation-dissipation theorem (FDT) obeyed by $U_{env}$. As can be seen from its expression in Eq.\eqref{eq:GreenRes}, $J(t)$ can be interpreted as the projection of the environment on an infinite set of harmonic oscillators\cite{Caldleg} and could be interpreted as a finite temperature Fourier transform. The harmonic oscillator is thus a cornerstone of the $P(E)$ theory and is worth a reminder as well. In the case of a single harmonic oscillator of frequency $\Omega$ at zero temperature, the probability $P(E)$ reads:
\begin{equation}
P(E)=\sum_{k\geq0}\frac{e^{-\rho}\rho^k}{k!}\delta(E-k\hbar\Omega).
\end{equation}
where $\rho=(e^2/2C)/h\Omega$ is the ratio between the charging energy of the capacitance and the energy of one mode. The consequences are two fold: First, the probability for the oscillator to absorb $n$ quanta obeys a Poissonian law that accounts for quanta emitted independently.  Second, the largest contribution to the current occurs for $k\approx \rho$. 

\begin{figure}
\vspace{-1cm}
\hspace{-.5cm}
\begin{pspicture}(0,0)(7,7)
\rput[bl](-.5,0){	\includegraphics[width=.45\textwidth]{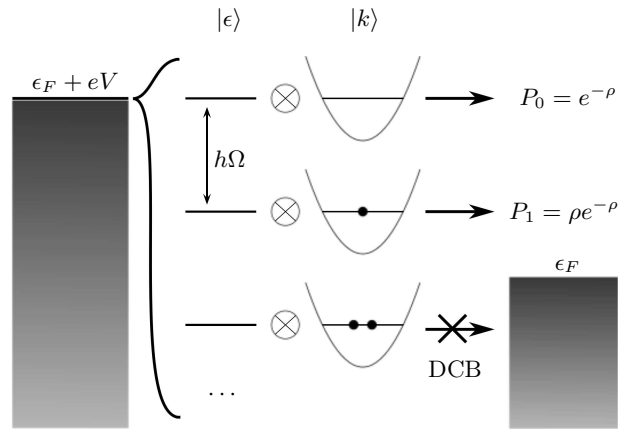}}
\rput[b](7,4.29){$P_0=e^{-\rho}$}
\rput[b](6.95,2.74){$P_1=\rho e^{-\rho}$}
\rput[b](5.5,0.7){DCB}
\rput[b](0.45,4.5){$\epsilon_F+eV$}
\rput[b](7.,2.1){$\epsilon_F$}
\rput[b](2.43,.4){$\cdots$}
\rput[b](2.2,3.0){\psline[linewidth=.75pt]{<->}(0,1.3)}
\rput[b](2.5,3.5){$h\Omega$}
\rput[b](2.5,5.3){$\ket{\epsilon}$}
\rput[b](4.3,5.3){$\ket{k}$}
\end{pspicture}
\caption{\label{DCBSketch} Sketch of the dynamical Coulomb blockade process. An electron of energy $eV$ coupled to a harmonic oscillator can be described by a superposition of states where the electron is of energy $eV-kh\Omega$ and the harmonic oscillator contains $k$ photons, $k\in\mathds N$. The probability of each state is given by a Poissonian law $P_k=e^{-\rho}\rho^k/k!$. }
\end{figure}

In order to provide a simple picture for this phenomenon,  let us consider a free electron described by a ket $\ket{\epsilon}$ coupled to a harmonic oscillator described by a ket $\ket{k}$, $k$ being the number of photons it holds. This new quantum system has its own eigenstates, which, at first order, can be written as the tensorial product of the eigenstates of both systems. The initial electronic state  can be written as a superposition of states $ \ket{\epsilon}\otimes\ket{0}\to\sum_k P_k\ket{\epsilon-kh\Omega}\otimes\ket{k}$. All these states have the same energy but have different weighs $P_k$ which are given by the Poissonian distribution $P_k=e^{-\rho}\rho^k/k!$ . This picture is sketched in Fig.\ref{DCBSketch}.
The $J(t)$ function\cite{IN} (see Eq.\eqref{eq:Jtfunction}) is the projection of the Green function of the environment on an infinite set of harmonic oscillators with a weigth $\Re [Z(\omega)]$  for each frequency. It is important to note that these harmonic oscillators are not damped and do not interact with each other. This is commonly justified by the fact that the system is linear.  However this is not the case in the well transmitting regime limit as we will see.


	\section{ P(E) theory for interacting electrons}

		\subsection{The Tunneling Limit}
Now we will extend $P(E)$ theory to the case where the leads on both sides of the tunneling junction are interacting. Notice that this situation corresponds also to the SBS regime, {\it i.e.} at low enough energy, when a weak impurity is inserted into an interacting wire. One dimensional gapless systems whose dispersion relation can be linearized at low energy can be described through their collective excitations, which are gapless bosonic modes. This description, called bosonization \cite{giama}, is done by introducing two bosonic conjugate fields: $\phi(x)$, which represents the charge of the TLL, and $\nabla\theta(x)$ its conjugate field such that $[\phi(x),\nabla\theta(x')]=-\ii\pi\delta(x-x')$.  The main advantage of this hydrodynamic description, is that Coulomb interactions can be treated exactly at low energy, giving rise to two parameters that characterize the systems: $K$ the interaction parameter, and $u$ the velocity of the collective charge excitations. In what follows, we mainly focus on repulsive Coulomb interaction $K<1$ ($K=1$ corresponds to a non-interacting system). In the FQHE, $K$ corresponds to the fractional filling at simple fractions $K=1/(2n+1)$, and the tunneling barrier is created by pinching off the 2D electron gas such that only electrons can tunnel between two edges.

Tunneling occurs between two distinct parts each described by a TLL Hamiltonian:
	\begin{equation}
		\HP_i=\frac{u}{2\pi}\displaystyle\int_{0}^{\infty} dx \left[K(\nabla\theta_i)^2+\frac{1}{K}(\nabla\phi_i)^2\right].
	\end{equation}
where $i$ denotes the left or the right of the barrier (see Fig.\ref{schemas}). The modes at $x\neq0$ can be integrated out as the action is quadratic and the tunneling term between the two TLL then reads\cite{kf}
	\begin{equation}
		\HP_T=\displaystyle v_T\cos 2\tilde\theta,
	\end{equation}
where $\tilde \theta=(\theta_R(t)-\theta_L(t))/2$. However, we need to couple the tunneling junction to the voltage drop $V$ through the junction and to the environment. Following
[\onlinecite{kf}],  this can be done by modifying the  tunneling hamiltonian $\HP_T$ as
	\begin{equation}
		\HP_T=v_T\cos(2\tilde\theta-\varphi+Vt).
	\end{equation}
with $\varphi(t)=e\int\limits_{-\infty}^t U_{\textrm{env}}(t') dt'$ the bosonic field  describing the environmental degrees of freedom  and $U_{\env}$ the potential at the edge of the environment. The tunneling current operator then reads:\cite{kf}
	\begin{equation}
		j_T=-e\frac{\updelta \mathcal H_T[\tilde\theta-\varphi+Vt]}{\updelta \tilde \theta }=2ev_T\sin 2(\tilde\theta-\varphi+Vt).
	\label{eq:jt}
	\end{equation}
In the next part we compute at lowest order in perturbation theory with respect to $v_T$ the average value of the tunneling current $\langle j_T\rangle$  and its correlations (the FF noise)
 using the electronic and environmental Keldysh Green functions defined in the appendix \ref{app:green}.

			\subsubsection{Average Current}
The main steps of the computation are given in App. \ref{app:back}. The tunneling current defined by $\mathcal I_T=\langle j_T\rangle$ then reads:
	\begin{equation}
\mathcal I_T
						= ev_T^2(1-e^{-\beta eV})\left[P_{}*\Gamma_K\right](eV),
	\label{eq:itune}
	\end{equation}
with
	\begin{equation}\label{eq:defpe}
		P_{}(E)=\int dt e^{-\ii Et}e^{-2\cor{(\varphi^{}(t)-\varphi^{}(0))^2}},
	\end{equation}
the probability for the environment to absorb an energy $E$ and 
 $\Gamma_K(E)$ reads
	\begin{equation}
		\Gamma_K(E)=\pi\beta\left(\tfrac{\pi\alpha}{\beta u}\right)^{\tfrac{2}{K}} B\left(\tfrac{1}{K}-\ii\tfrac{\beta E}{2\pi},\tfrac{1}{K}+\ii\tfrac{\beta E}{\pi}\right)e^{\tfrac{\beta E}{2}}.\\
	\label{eq:LNK}
	\end{equation}
We have introduced $\alpha^{-1}$ as a high momentum cut-off and $B(a,b)=\Gamma(a)\Gamma(b)/\Gamma(a+b)$ the Beta function. By analogy with Eq. \eqref{eq:PET}, the $\Gamma_K^{^{}}(E)$ can be interpreted as the probability for an electron to tunnel through the tunnel barrier.\cite{Wen} It is worth pointing that Eq. \eqref{eq:itune} constitutes an extension of the P(E) theory to an interacting wire. We note that Eq. \eqref{eq:PET} is recovered by setting $K=1$. 

			\subsubsection{Finite frequency noise}
The non-symmetrised noise at finite frequency is given by the Fourier transform of:
	\begin{equation}
		S_T(t)=\cor{j_T^{}(t)j_T^{}(0)}-\mathcal I_T^2.
	\end{equation} 
The second term is in fourth order in $v_T$ and can be neglected. 
 Details of the calculation can be found in the appendix \ref{app:back}. One obtains at second order in $v_T$:
	\begin{equation}
	S_T(\omega)
	=e\sum_{\eta=\pm} {\cal N}((\omega+\eta eV))\mathcal I_T(\omega+\eta eV), \label{eq:DFTTu}	
	\end{equation}
with ${\cal N}$ the Bose distribution. 
The last expression is an out-of-equilibrium and perturbative fluctuation dissipation theorem (FDT). 
Actually,  this out-of-equilibrium perturbative FDT can be derived for arbitrary tunnel junctions (between conductors of arbitrary dimensions, with internal and mutual interactions, and coupled to electromagnetic environment or quantum systems at equilibrium).\cite{Safi13} Note also that the same relation has been obtained in a TLL with a tunneling barrier, thus to a conductor coupled to an ohmic environment.\cite{BenaSafi} A similar but different relation is obeyed as well by the {\it symmetrized} FF noise, and has been derived by specifying either to an interacting conductor without an environment,\cite{zaikin} or to a noninteracting conductor in series to an external low impedance environment.\cite{Levitov,yeyati} 
Going back to  Eq. \eqref{eq:itune}, and taking its derivative with respect to voltage, one gets
	\begin{equation}\label{eq:fdtT}
		G_T(eV)=\iinfty d\omega  P(\omega) \partial_\omega S_T(-\omega).
	\end{equation}	
We hence recover a generalisation of [\onlinecite{Levitov}] and [\onlinecite{Gab08}] that relates directly the variation of the differential conductance to the shot noise in the absence of an environment. 

		\subsection{The WBS limit}
\begin{figure}
		\vspace{-1.5cm}
		\begin{pspicture}(0,0)(7,8.2)
		\rput[bl](-.5,0.3){\includegraphics[width=0.45\textwidth]{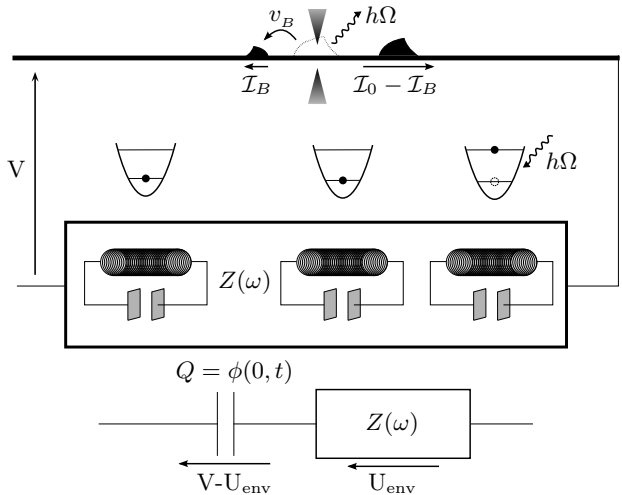}}
		\rput[b](4.4,6.25){$h\Omega$}
		\rput[b](6.8,4.3){$h\Omega$}
		\rput[b](4.6,5.3){$\mathcal I_0-\mathcal I_B$}
		\rput[b](2.75,5.3){$\mathcal I_B$}
		\rput[b](3.1,6.2){$v_{_B}$}
		\rput[b](4.55,0.75){$Z(\omega)$}
		\rput[b](2.59,2.65){$Z(\omega)$}
		\rput[b](2.45,0){V-U$_{\textrm{env}}$}
		\rput[b](4.55,0){U$_{\textrm{env}}$}
		\rput[b](2.45,1.45){$Q=\phi(0,t)$}
		\rput[b](-.2,2.9){\psline[linewidth=.6pt]{->}(0,2.7)}
		\rput[b](-.45,4.2){V}
		\end{pspicture}
	\caption{\label{schemasbs}On top, a "QPC" coupled to an environment (the WBS limit). As a the "QPC" acts on the charge density wave, part of it is reflected and some of its energy is absorbed by the environment. Bottom, equivalent circuit from the point of view of the "QPC".}
	\end{figure}

Now we consider a  weak impurity embedded into a TLL with parameter $K$. The situation is modeled on Fig. \ref{schemasbs}. In the case of FQHE, this corresponds to a quantum point contact where fractional charges $Ke$ can tunnel between the two chiral edges.
	
	The Hamiltonian for the TLL is given by:
	\begin{equation}
		\HP_0=\frac{u}{2\pi}\displaystyle\int_{-\infty}^{\infty} dx \left[K(\nabla\theta)^2+\frac{1}{K}(\nabla\phi)^2\right].
	\end{equation}

The impurity is modelled by a local scattering potential. As shown in [\onlinecite{kf}], the backscattering Hamiltonian, in a bosonized language, reads:
	\begin{equation}
		\mathcal H _B=\sum_n v_B^{2n}\cos(2n\phi(x=0)),
	\end{equation}
	where $v_B$ denotes the backscattering amplitude.
The local character of the scattering potential ensures that backscattering events are instantaneous.
Also, multi-electronic processes $(n>1)$ can be ignored when $Kn^2>1$.\cite{kf}  We  neglect forward scattering which does not affect here current voltage characteristics.\cite{tztl2} As before, we can integrate out the fields at $x\neq0$.
 The TLL is driven out of equilibrium and the voltage drop is assumed to be at the QPC. \cite{zaikin,Vdrop} The charge at the QPC is then coupled to the electric fields induced by the bias and the environment:\cite{Safi}
	\begin{equation}
		\HP_V=\frac{1}{\pi}\partial_t(A(t)-\varphi)\phi(x=0),
	\label{eq:HPV}
	\end{equation}
where $A(t)$ is the potential vector defined by $\partial_tA(t)=V(t)$.

			\subsubsection{Average current}
For a quantum wire undergoing a voltage, the current can be separated into two contributions:
	\begin{equation}
		\mathcal I=\mathcal I_0-\mathcal I_B,
	\end{equation}
where $\mathcal I_0$ is the  current for a clean wire, and the backscattering current $\mathcal I_B$ is due to backscattering. These quantities can be computed from the TLL Green functions.\cite{tztl} Here, since the electronic liquid and the electromagnetic environment are linearly coupled, we can integrate out either the electronic modes or the environmental modes leading to the effective dressed propagators:
	\begin{eqnarray}
		&\CP_{\eff}^{-1}(\omega_n)&=\CP_{\el}^{-1}(\omega_n)+\left(\frac{\omega_n}{\pi}\right)^2\CP_{\env}(\omega_n),\label{eq:Ceffel}\\
  	&\CP_{\bos}^{-1}(\omega_n)&=\CP_{\env}^{-1}(\omega_n)+\left(\frac{\omega_n}{\pi}\right)^2\CP_{\el}(\omega_n),\label{eq:Ceffb},
	\end{eqnarray}
where $\CP_{\eff}$ and $\CP_{\bos}$ are the effective Green functions of the electrons and of the environment respectively. They are defined and calculated in the appendix
\ref{app:gosc}. As perturbative computations lead us to compute them at equilibrium, in absence of a barrier, one can use the Matsubara formalism for that, before switching to real time. 
By combining the two equations, one may eventually write $\CP_{\eff}$ in terms of $\CP_{\bos}$ and $\CP_{\el}$:
\begin{equation}
\CP_{\eff}(\omega_n) =\CP_{\el}(\omega_n)-\left(\frac{\omega_n}{\pi}\right)^2\CP_{\bos}(\omega_n)\CP_{\el}(\omega_n)^2,
\end{equation}
the effective action for the environment can be immediately deduced by inverting the subscripts. In the WBS limit, the back-action of the environment cannot always be neglected and the consequences are numerous.
The voltage variations at the edge of the QPC and of the environment affect each other. For the QPC, the corresponding potential vector reads, for a DC voltage:
	\begin{equation}
		A(t)=KeVt-\frac{K^2\pi eV}{2\ii}\int_0^t \CP_{\bos}^R(t')dt' .
	\label{eq:Apot}
	\end{equation}
The backscattering current operator is defined by:\cite{tztl}
	\begin{equation}
		j_B=-e\frac{\updelta \HP_B[\phi]}{\updelta \phi}.
	\label{eq:jback}
	\end{equation} 
At second order in perturbation with respect to $v_B$, its average reads:
	\begin{equation}
		\mathcal I_B(t)=\langle j_B\rangle= \frac{2}{\pi}\iinfty dt' \partial_t\CP_{\eff}^R(t-t')\cor{j_B}(t').
	\end{equation}
In the general case, $\cor{j_B}$ is time-dependent. 
 In order to be as general as possible, we separate the linear and the non-linear parts, with respect to time, of the potential vector:
	\begin{equation}
		A(t)=K'eVt+f(t),
	\label{eq:assumption}
	\end{equation}
where $K'$ is an effective local Luttinger parameter that embodies the dissipative behaviour of the environment.\cite{Safi} Let us first assume that $|f(t)|\ll K'eV|t|$ so that $A(t)$ is linear in time. Consequently, the backscattering current $\langle j_B\rangle$ becomes time-independent. 
The general case will be discussed in Sec. \ref{sec:ac}.
Under this approximation, the average backscattering current can be computed (see appendix \ref{app:back} for details) and reads 
	\begin{equation}
		\mathcal I_B=\displaystyle e^*v_B^2(1-e^{-\beta e^*V})\left[P^{^{}}_{}* \Gamma^{^{}}_{1/K'}\right](e^*V), 	
	\label{eq:itrans}
	\end{equation}
	where $e^*=K'e$ is the effective charge which is scattered, and $P^{^{}}_{}$ is defined in Eq. (\ref{eq:defpe}). 
We recover again a duality between the WBS and the SBS regimes: the transition rate $\Gamma_K^{\eta\eta'}$ can be obtained for both conducting regimes by changing $K\to1/K$. Consequently, the current follows a $V^{2K-1}$ power law, meaning that the backscattered current is depleted by electronic interactions.
 The main difference lies in the nature of the scattered particle, in the transparent regime it is a charge density wave of effective charge $e^*$ that is scattered, while this is an electron of charge $e$ which tunnels through the potential barrier. Yet, it is important to note that in experiments involving a quantum wire, due to the nature of the contacts with the measuring device or with the reservoirs which behave like Fermi liquids, the charge of the excitation remains $e$.\cite{tztl2} The previous results remain correct by setting $e^*=e$ and assuming that all energy scales are larger than  $u/L$ where $L$ is the length of the wire.

			\subsubsection{Finite frequency noise}
Following [\onlinecite{tztl}] and [\onlinecite{BenaSafi}], and  using the same framework that allowed us to compute Eq. \eqref{eq:itrans}, we can write 
the FF non-symmetrized noise  as the sum of three contributions:
	\begin{equation}
		S_{{\rm WBS}}(\omega)=S_0(\omega)+S_B(\omega)+S_C(\omega),
	\end{equation}
$S_0$ is the FF noise at $v_B=0$, $S_B$  the shot noise and $S_C$ can be thought of as a kind of cross correlator.
Positive  frequencies and negative frequencies correspond to emission and absorption of photons respectively.

 Using the Keldysh Green functions $C^K_{\eff}(t)$, we obtain the following expressions for the three noise components:
	\begin{equation}
S_0(\omega)=2\displaystyle (\tfrac{e\omega}{\pi})^2 {\cal N}(\omega)\left(\CP_{\eff}^R(\omega)-\CP_{\eff}^A(\omega)\right),
\end{equation}	
the Johnson Nyquist noise in absence of backscattering, 
\begin{equation}		\label{eq:sagen}
S_B(\omega)=-\dfrac{2e}{\pi}\CP_{\eff}^A(\omega)\CP_{\eff}^R(\omega)
\displaystyle \sum_{\eta=\pm}\mathcal{I}_B(\omega+\eta eV){\cal N}(\omega+\eta eV)
\end{equation}
the FF shot noise
and finally
\begin{eqnarray}
		&&S_C(\omega)=2\left(\tfrac{e\omega}{\pi}\right)^2{\cal N}(\omega)(\CP_{\eff}^A(\omega)-\CP_{\eff}^R(\omega))\displaystyle\int dt~ \sgn(t) \nonumber\\
		&& \times \cos(eVt)e^{2\GP (t)}\displaystyle\left(\CP_{\eff}^A(\omega)e^{\ii\omega|t|}+\CP_{\eff}^R(\omega)e^{-\ii\omega|t|}\right),
		\label{eq:sc}
	\end{eqnarray}
a cross correlator. $\GP$ denotes the charge field Green function (see Eq. \ref{eq:defG}).
In order to derive these results, we have again assumed that $A(t)$ is linear in time.

 Eq. (\ref{eq:sagen}) relates the FF shot noise $S_B$  to the backscattering current $\mathcal{I}_B$. 
 Such relation is the equivalent of Eq. (\ref{eq:DFTTu}) derived for the tunneling regime, and more universally in Ref.\onlinecite{Safi13}.

If the advanced Green function is odd with respect to $\omega$, the sum  $S_B+S_C$ reduces to:
\begin{equation}
\begin{array}{c}
(S_A+S_C)(\omega)=-\dfrac{2e}{\pi}\CP_{\eff}^A(\omega)\CP_{\eff}^R(\omega)\times \\
\displaystyle \sum_{\eta=\pm}\mathcal{I}_B(\omega+\eta eV)\left({\cal N}(\omega+\eta eV)-2{\cal N}(\omega)\right).
\end{array}
\label{eq:scsimply}
\end{equation}
The term proportional to ${\cal N}(\omega\pm eV)$ in Eq. (\ref{eq:scsimply}) accounts for the non-symmetrized shot noise $S_B$ while the term in ${\cal}(\omega)$ accounts for $S_C$. These two terms have an opposite sign and are in competition. The production of backscattering current enhances the noise but the QPC also affects  it through the cross term. We will see consequences of this competition in Sec. \ref{sec:ffn} when we will treat the case of the LC resonator as the environment.

By taking the derivative of the backscattering current $\mathcal I_B$ defined in Eq. (\ref{eq:itrans})  with respect to the bias, one gets the backscattering differential conductance $G_B=\frac{d\mathcal I_B}{dV}$:
	\begin{equation}\label{eq:ga}
		G_B(eV)=\iinfty d\omega P_{}(\omega) \partial_\omega S_B(-\omega).
	\end{equation}
Note that the total conductance is defined by $G=G_0-G_B$.
As for the tunneling regime, we therefore find a relation between the differential backscattering conductance in presence of an environment and the shot noise $S_B$ in the absence of an environment. Eq. (\ref{eq:ga}) can be regarded as the equivalent of Eq. (\ref{eq:fdtT}) in the WBS regime. However, we want to stress that contrary to the tunneling regime, Eq. (\ref{eq:ga}) only involves a part of the total FF noise.
 

\section{Transport through a QPC coupled to various environments}
\begin{figure}[t]
\begin{pspicture}(0,0)(7,6)
\rput[bl](-.5,0){\includegraphics[width=0.45\textwidth]{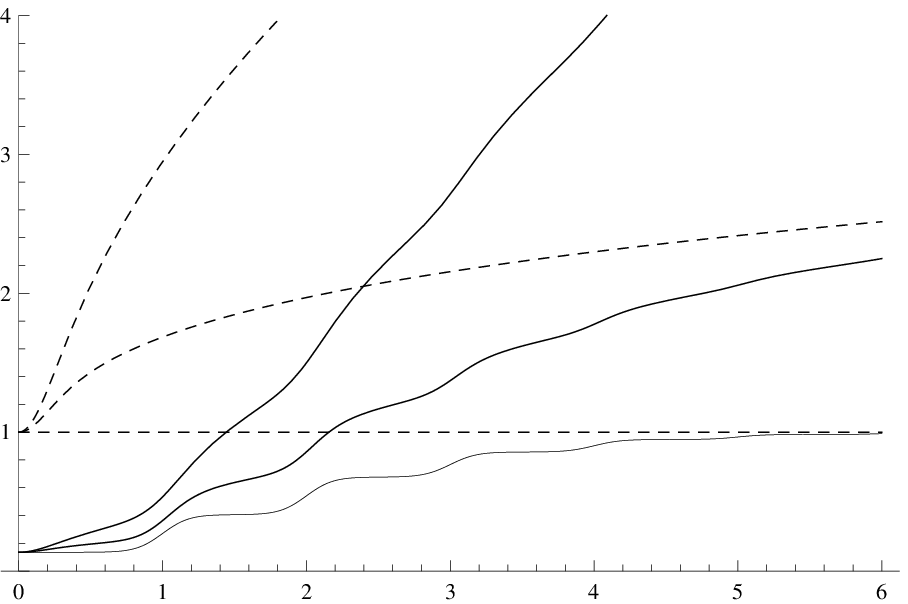}}
\rput[b](7,.45){$eV/h\Omega$}
\rput[b](6.9,3.5){$K=0.8$}
\rput[b](6.8,1.7){$K=1.0$}
\rput[b](5.5,5){$K=0.6$}
\rput[bl](-.3,5.3){$G_T(eV)/\left.G_T(0)\right|_{\rho=0}$}
\end{pspicture}		
	\caption{	\label{Tunnel_Cond_BCD} Ratio between the non-equilibrium differential tunneling conductance  and its zero-bias value $\left.G_T(0)\right|_{\rho=0}$ (with no environment) with (full lines) and without (dashed lines) an environment as a function of $eV/h\Omega$. We fix  $\rho=1$. This plot is done for three different values of the Luttinger parameter: $K=1$, $K=0.8$ and $K=0.6$.} 
\end{figure}
\begin{figure}[t]
\begin{pspicture}(0,0)(7,6)
\rput[bl](-.5,0){\includegraphics[width=0.45\textwidth]{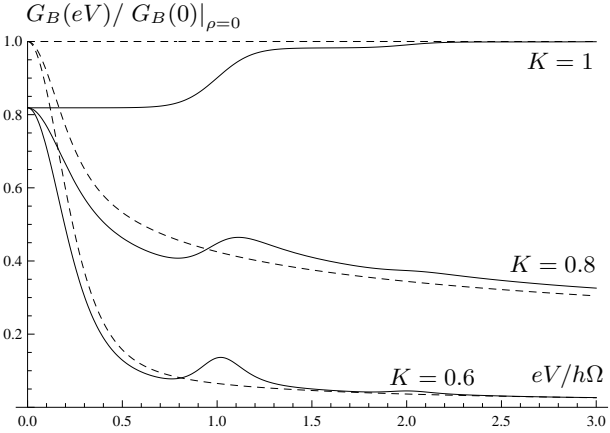}}
\rput[b](7,.55){$eV/h\Omega$}
\rput[b](6.9,4.8){$K=1$}
\rput[b](6.8,2.1){$K=0.8$}
\rput[b](5.2,.6){$K=0.6$}
\rput[bl](-.2,5.3){$G_B(eV)/\left.G_B(0)\right|_{\rho=0}$}
\end{pspicture}		
	\caption{\label{Trans_Cond_BCD} Ratio between the differential backscattering conductance and its zero-bias value $G_B^0(0)$ (with no environment) with (full lines) and without (dashed lines) an environment  as a function of $eV/h\Omega$. We  fixed  $\rho'=0.1$ and took three different values for the Luttinger parameter: $K=1$, $K=0.8$ and $K=0.6$.} 
\end{figure}

In this section we apply the previous formalism to analyze the transport properties of a QPC in series with two paradigmatic environments: the harmonic oscillator which may describe a resonator such as a LC circuit and the ohmic environment. The  feedback regime of an environment can be studied in both transmitting regimes.

\subsection{The harmonic oscillator}

Here we study the influence of a harmonic oscillator on the current in both transmitting regimes. Our basic ingredients are  $\CP_{\el}(t)$ the Green function for the interacting electrons and $\CP_{\res}$ the Green function for the oscillator. They are defined and computed in the appendix. 
The back-action cannot be neglected as in the transparent limit without taking some care. When  the electrons flowing through the QPC are coupled to the environment, it is convenient to work with $\CP_{\bos}$, the environmental effective Green function introduced in Eq. (\ref{eq:Ceffb}) which reads for the case of a harmonic oscillator:
\begin{equation}
\CP_{\bos}(\omega_n)=\frac{\pi}{R_KC}\frac{1}{\omega_n^2+2|\omega_n|/\tau_R+\Omega^2},
\end{equation}
where $\tau_R=4\pi R_KC/K$ is the relaxation time of the damped harmonic oscillator.\cite{Caldleg,Schmid} For $(K/2)^2\rho\ll1$, using Eq. (\ref{eq:Ceffel}), and switching to real time, we can approximate the effective Green function of the TLL by:
\begin{equation}\label{eq:geffe}
\CP_{\eff}(t)=\CP_{\el}(t)-\left(\frac{K}{2}\right)^2\CP_{\res}(t).
\end{equation}
The potential vector is written as:
\begin{equation}
A(t)=KeVt\left(1-\frac{K\rho}{4\pi\Omega t}[\cos(\Omega t)-1]\right),
\end{equation}
so that the ac contribution can be neglected. The effective electronic Green function in Eq. (\ref{eq:geffe}) is then the sum of the TLL Green function and of the harmonic oscillator Green function with a renormalized $\rho'=\rho (K/2)^2$ parameter. The bare current reads:
\begin{equation}
\mathcal I_0(t)=\tfrac{2\ii}{\pi} G_0V\int\partial_t\CP_{\eff}^{R}=G_0V\left(K-\rho'\sin(\Omega t)\right).
\end{equation}
Again, owing to the harmonic oscillator, an ac component of the voltage arises.

\subsubsection{Average current}
We are now left for both  SBS and WBS regimes with the dual equations for the tunneling and the weak backscattering current.
\begin{equation}
\begin{array}{c}
\mathcal I^0_T=\tfrac{ev_T^2}{\pi}e^{-\rho\coth(\tfrac{\beta h\Omega}{2})}\displaystyle\sum_{k=-\infty}^{\infty}P_k(\rho)\\
\times \left(\Gamma_{K}(-kh\Omega-eV)-\Gamma_{K}(eV-kh\Omega)\right),
\end{array}
\label{eq:itunosc}
\end{equation}
\begin{equation}
\begin{array}{c}
\mathcal I^0_B=Kev_B^2e^{-\rho'\coth(\tfrac{\beta h\Omega}{2})}\displaystyle\sum_{k=-\infty}^{\infty}P_k(\rho')\\
\times\left(\Gamma_{1/K}(e^*V-kh\Omega)-\Gamma_{1/K}(-e^*V-kh\Omega)\right),
\end{array}
\label{eq:itransosc}
\end{equation}
where
\begin{equation}
P_k(\rho)=I_k\left(\frac{\rho}{\sinh(\tfrac{\beta h\Omega}{2})}\right)e^{\tfrac{k\beta h\Omega}{2}},
\end{equation}
is the finite temperature version of the Poissonian law. The probability $\Gamma_K^{}$ for an electron to tunnel is non zero when its argument is positive, so that there is a new contribution of amplitude $P_k(\rho)$ to the current every time $|e^*V|>kh\Omega$. Physically, this is equivalent to the reopening of a conducting channel by the environment which was dynamically blocked before, leading to the contributions of the second higher mode to the current. As we sum up all the modes $k$ we end up with the effective differential conductance of the system which is plotted in Fig. \ref{Tunnel_Cond_BCD} for the tunnel regime and on Fig. \ref{Trans_Cond_BCD} for the WBS regime.
\begin{figure}
\begin{pspicture}(0,0)(7,5.5)
\rput[bl](-.7,0){	\includegraphics[width=.45\textwidth]{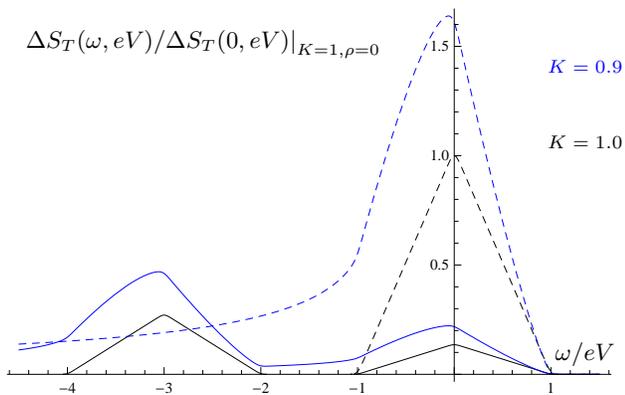}}
\rput[b](7.1,0.4){$\omega/eV$}
\rput[b](7.1,4.3){\scriptsize{\color{blue}{$K=0.9$}}}
\rput[b](7.1,3.3){\scriptsize {\color{black}{$K=1.0$}}}
\rput[b](2,4.5){$\left.\Delta S_T(\omega,eV)/\Delta S_T(0,eV)\right|_{K=1,\rho=0}$}
\end{pspicture}
\caption{\label{FENoiseTu} (Color Online) Ratio between the differential finite frequency excess noise of the tunnel current and its value at $\omega=0$, $K=1$ and $\rho=0$ as a function of $\omega/eV$ with (full line) and without (dashed line) a coupling to the harmonic oscillator. This plot is done for a finite bias $h\Omega/eV=3$, $\beta h\Omega=16.5$,  and $\rho=2$. and for $K=1$(black), $K=0.9$ (blue).}
\end{figure}
Without coupling to the environment, the differential conductance  behaves   as  the power law  $G_T(eV)\sim G_T(0)(eV/h\Omega)^{\frac{2}{K}-2}$ in the tunneling regime
(see  Fig. \ref{Tunnel_Cond_BCD}) and as $G_B(eV)\sim G_B(0)(eV/h\Omega)^{2K-2}$ in the WBS regime (see Fig. \ref{Trans_Cond_BCD}).
When the TLL is coupled to the harmonic oscillator, the differential conductance tends asymptotically toward its value but with steps at voltage $eV=h\Omega$ or $e^*V=h\Omega$.
 
																		\subsubsection{Finite Frequency Noise}\label{sec:ffn}

For strongly interacting systems, the singularity at $eV=kh\Omega$ which could be thought of as a hallmark of the contribution of the k$^{\textrm{th}}$ mode of the harmonic oscillator is generically smeared out by interactions. 
In order to highlight the role of the harmonic oscillator, we analyse the non-symmetrized excess shot noise at finite frequency in both cases\cite{BenaSafi,Bena} which is defined as:
\begin{equation}
\Delta S_{T,B}(\omega,eV)=S_{T,B}(\omega,eV)-S_{T,B}(\omega,0).
\label{eq:Sexcess}
\end{equation}

We first focus on the tunneling FF noise (Fig. \ref{FENoiseTu}). For 
non-interacting electrons and with no environment (dashed black curve), this quantity is, as expected, linear, positive, even in frequency,  and vanishes for $h|\omega|>|eV|$. 
As we couple the non-interacting electrons to the harmonic oscillator (plain curves), the triangular-shape pattern  of the excess noise around $\omega=0$ is strongly decreased and a new identical pattern arises around $\omega=-\Omega$ breaking the emission absorption symmetry, a fact attributed universallly to non-linear transport.\cite{SafiJoyez} Going back to Fig.\ref{DCBSketch}, we see that this pattern accounts for the absorption of  photons of energy of order $h\Omega$ by the oscillator and its height is directly proportional to $P_1$, the probability for an electron to be in the state $\ket{eV-h\Omega}\otimes\ket{1}$. The total area underneath each curve is not affected by the harmonic oscillator as the system gives back all the energy that is given to it when driven out of equilibrium. 
In presence of interactions, but in absence of the environment, we recover the FF noise obtained in a TLL with a weak impurity \cite{BenaSafi}, where the emission-absorption symmetry is broken as well due to non-linearity. When connected to an LC resonator, and due to interactions, the electronic conductor can absorb more photons when polarized, even at $\omega<-eV$ as can be seen on the plain blue curve. We consider moderate interactions ($K=0.9$) to draw the plots in Fig. \ref{FENoiseTu} in order to visualize on the same plot the interacting and the non-interacting case.

\begin{figure}[]
\begin{pspicture}(0,0)(7,5.5)
\rput[bl](-.7,0){	\includegraphics[width=.45\textwidth]{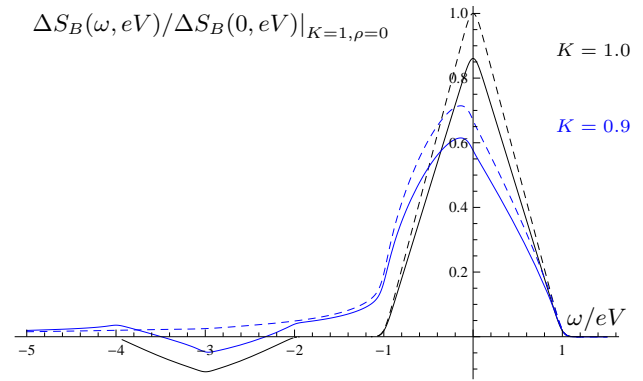}}
\rput[b](7.15,0.65){$\omega/eV$}
\rput[b](7.1,3.3){\scriptsize{\color{blue}{$K=0.9$}}} 
\rput[b](7.1,4.3){\scriptsize {\color{black}{$K=1.0$}}}
\rput[b](2,4.5){$\left.\Delta S_B(\omega,eV)/\Delta S_B(0,eV)\right|_{K=1,\rho=0}$}
\end{pspicture}
\caption{\label{FENoiseTr} (Colour Online) Ratio between the differential finite frequency excess noise in the WBS limit and its value at $\omega=0$, $K=1$ and $\rho=0$, as a function of $\omega/eV$ with (full line) and without (dashed line) a coupling to the harmonic oscillator. This plot is done for a finite bias $h\Omega/eV=3$, $\beta h\Omega=16.5$,  and $\rho=0.15$. and for $K=1$(black), $K=0.9$ (blue).}
\end{figure}

In the WBS limit (see Fig. \ref{FENoiseTr}), the physics is much richer as we have two contributions to the FF noise: $S_B$ which has identical features as the tunneling shot noise and $S_C$ the cross correlator. 
The bare noise $S_0$  does not contribute to the excess noise. 
We can directly use Eq. (\ref{eq:scsimply}) to plot  the FF noise in Fig. \ref{FENoiseTr}.
In the non-interacting case, and when no environment is coupled to the system, the $S_C$ term is voltage independent
and does not contribute to the excess noise. 
We thus recover the usual triangular shape (black-dashed curve). As a non-interacting wire is coupled to a harmonic oscillator (full black  curve), we see that a negative triangle appears around $\omega=-\Omega$ suggesting that less noise is absorbed by the electrons when the system is undergoing a bias. More generally, it is possible to obtain a negative excess FF noise only in the absorption spectrum, and for nonlinear systems,\cite{SafiJoyez} which is the case here.  This is consistent with the picture given in Fig. \ref{DCBSketch}. At small temperatures ($\beta \omega\gg1$), the emitted noise can only be attributed to $S_A$. As in the tunnel limit, the density of states is affected by the oscillator and part of the electrons are in a $\ket{eV-h\Omega}\otimes\ket{1}$ state. Based on $S_A$, we should therefore have expected a noise enhancement at $\omega =eV/h-\Omega$. However, the $S_C$ contribution is active and has an opposite sign and ultimately leads to a  reduction of the absorption noise. 
Contrary to the tunneling case, 
the total area under the curve is not conserved in the WBS regime.

When interactions are switched on, the triangles become more rounded  (blue curve) but the physics remains qualitatively the same.

\subsubsection{Role of feedback}\label{sec:ac}
\begin{figure}[t]
\begin{pspicture}(0,0)(7,6)
\rput[bl](-.5,0){\includegraphics[width=0.45\textwidth]{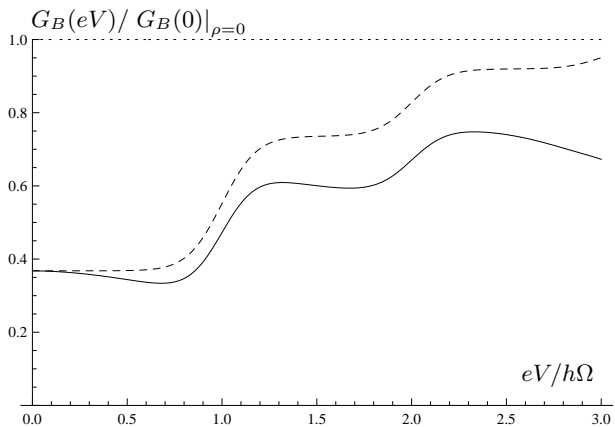}}
\rput[b](6.8,.55){$eV/h\Omega$}
\rput[bl](-.2,5.2){$G_B(eV)/\left.G_B(0)\right|_{\rho=0}$}
\end{pspicture}		
	\caption{\label{Trans_Cond_BCD_ac} Ratio between the differential backscattering conductance and its zero-bias value $G_B^0(0)$ (with no environment) with (full lines) and without (dashed and dotted lines ) taking potential fluctuations into account as a function of $eV/h\Omega$. We  fixed  $\rho'=0$ (dotted line) and $\rho'=1$ (dashed and full line) } 
\end{figure}

In the previous part, we have neglected the time-dependent part of the potential vector in Eq. (\ref{eq:assumption}) and the feedback of the TLL on the oscillator, bounding us to a $\rho\ll1$ regime. The aim of this paragraph is to understand, at least qualitatively, at first order in $\rho$, how the time-dependant part of the potential vector affects the current. 
First, we assume that the essential effect of the TLL is to damp the harmonic oscillator. This corresponds to a Markovian approximation in which the information held by the environment (here the oscillator) on the history of the system can be neglected. The harmonic oscillator acquires a finite lifetime $\tau_R$ which enables us to compute its retarded Green function:
\begin{equation}
\partial_tC_{\eff}^R(t)\approx\delta(t)\CP_{el}^{K}+\theta(t)\rho\cos(\Omega t),
\end{equation}
and hence the potential at the edge of the QPC:
\begin{equation}
f(t)=\frac{eV\rho'}{\pi^2\tilde \Omega}\left(1 -e^{\frac{-t}{\tau_R}}\cos(\tilde\upsilon t)\right),
\label{eq:potential}
\end{equation}
from which we obtain:
\begin{equation}
\mathcal I_B=K\cor{j_B}(0)+\iinfty \theta(t)\rho'\cos(\Omega t)j_B(t-t').
\end{equation}
At first order, $j_B$ is a constant and the second term can be neglected. For small damping, Eq. \eqref{eq:itune} becomes:
\begin{widetext}
\begin{equation}
\begin{array}{rl}
\mathcal I_B(eV)&=J_0\left(\tfrac{eV\rho'}{\pi^2\tilde\Omega}\right)\left(\cos(\tfrac{eV\rho'}{\pi^2\tilde\Omega})\mathcal I_B^0(eV)-\sin(\tfrac{eV\rho'}{\pi^2\tilde\Omega})\mathcal I_B^1(eV)\right)\\
&+\displaystyle\sum_{k=1}^{\infty}(-1)^kJ_{2k}(\tfrac{eV\rho'}{\pi^2\tilde\Omega})\left(
\cos(\tfrac{eV\rho'}{\pi^2\tilde\Omega})\sum_{\eta=\pm} \mathcal I_B^0(eV+ 2\eta kh\tilde\Omega)
-\sin(\tfrac{eV\rho'}{\pi^2\tilde\Omega})\sum_{\eta=\pm} \mathcal I_B^1(eV+ 2\eta kh\tilde\Omega)\right)\\
&+\displaystyle\sum_{k=0}^{\infty}(-1)^{k}J_{2k+1}(\tfrac{eV\rho'}{\pi^2\tilde\Omega})\left(
\sin(\tfrac{eV\rho'}{\pi^2\tilde\Omega})\sum_{\eta=\pm} \mathcal I_B^0(eV+ (2k+1)\eta h\tilde\Omega)
-\cos(\tfrac{eV\rho'}{\pi^2\tilde\Omega})\sum_{\eta=\pm} \mathcal I_B^1(eV+\eta (2k+1)h\tilde\Omega)
\right),
\end{array}
\label{eq:final}
\end{equation}
where $J_k$ is the k$^\mathrm{th}$ Bessel function, and 
\begin{equation}
\mathcal I^1_B(E)=e^*v_B^2\sum_\eta\displaystyle\int dt \cos(E t)e^{-2\GP^{\eta\eta}(t)}=\tfrac{2^{^{2K}}e^*v_B^2\beta}{2\pi} \sin(K\pi)\displaystyle\sum_{n\in\mathds Z,\eta}P_nB\left(1-2K,1-K+\frac{\ii\beta}{2\pi}( nh \tilde\Omega+\eta E)\right).
\label{eq:IB1}
\end{equation}
\end{widetext}
From this equation, two main types of contributions
can be extracted. First, we recover the main current we computed before, but which is now modulated by a
Bessel function ($J_0$). In addition to that, a term that involves the emission or absorption of photons emerges. It can be understood as the result of the interactions 
between the backscattering current and the photons inside the oscillator which were emitted by the bare current, leading either to photo-assisted or dynamically blocked currents, depending on the voltage and on the strength of the interactions.
Note that the current $I_B^1$ in Eq. (\ref{eq:IB1}) is non-zero only when interactions are taken into account.
 The corrections due to the fluctuating potential are plotted in Fig. \ref{Trans_Cond_BCD_ac}.

\subsection{The ohmic environment}
We have presented in the previous section a detailed study of the QPC in series with a resonator. Let us comment in this paragraph on the case of the ohmic resistance.
Following Ref. [\onlinecite{Caldleg}], the harmonic oscillator can be regarded as the buiding block to design any arbitrary environment. In particular, the Ohmic environment can be decomposed
as an infinite set of harmonic oscillators. We therefore consider a QPC coupled to a resistance and particularly focus on  the WBS limit, the most interesting situation.
In this case, Eq. \eqref{eq:Ceffel} becomes:
\begin{eqnarray}
&\CP^{-1}_{\eff}(\omega_n)&=\CP_{\el}^{-1}(\omega_n)+\omega_n^2\sum_k \frac{\pi}{R_KC}\frac{1}{\omega_n^2+\Omega_k^2}\\
&&=\frac{2}{\pi}|\omega_n|\left(\frac{1}{K}+\frac{R}{R_K}\right),
\end{eqnarray}
where R is the equivalent resistance of the set of harmonic oscillator. It appears clearly that at the QPC, the system behaves locally like a TLL of parameter $K'=K/(1+KR/R_K)$. The same result was obtained in Ref. [\onlinecite{Safi}] in both limits. A direct consequence is that the effective potential vector is linear and the Markovian assumption can be relaxed. This can be understood by the fact that the retarded Green function is a constant, in other words that the system is purely dissipative and the environment discards any information it has on the TLL and reciprocally.
Therefore, in presence of an ohmic environment, the current and the FF non-symmetrized noise correspond to the results obtained in a TLL with an interaction parameter $K'$.\cite{BenaSafi}

\section{Conclusion}
In this paper we have studied the non-equilibrium transport properties of a TLL with an interaction parameter $K$ and a weak  or strong  backscattering center coupled to an electromagnetic environment. The SBS regime corresponds also to a tunneling barrier between two TLLs.  
Our results apply to a noninteracting conductor, or a QPC, by letting $K=1$, or a conductor connected to a resistance $R$, Eq.(\ref{key}). We have obtained formal expressions of  the non-symmetrized FF noise, as function of bias and temperature, similar to those obtained in a TLL without an environment,  provided one uses the renomalized bosonic Green functions. This can be understood by the equivalence between a conductor connected to an arbitrary impedance $Z(\omega)$ and a one dimensional wire with finite--range interactions \cite{jezouin} for which the Keldysh approach developed in TLLs as well as the exact formal expressions for the FF noise in Refs.\onlinecite{Bena} are appropriate as well.     
Then we have derived perturbative expressions with respect to either the weak or tunneling amplitudes of the barrier. In particular, we have obtained again strong deviations from duality between the SBS and WBS regimes, which hold not only in the form of the FF noise expressions, but also in the back action of the environment. In the tunneling (thus SBS) regime, general expressions for  transport quantities extending the well-established P(E) theory to include electronic interactions have been obtained. Moreover, we have also found general relations between the differential conductance in the presence of an environment to the non-symmetrized FF shot noise without the environment for both the SBS regime (see Eq. (\ref{eq:fdtT})) and the WBS regime (see Eq. \ref{eq:ga})).

We have then computed explicitly the  non linear DC conductance and the FF noise when the TLL is in series with a harmonic oscillator, such as a LC resonator, in the limit of weak  coupling. The FF noise has been shown to have two features: asymmetry between  the emission and absorption spectrum, and domains of (negative) frequencies for which the excess (absorption) noise can be negative; both are consequences of the nonlinearity of the system.\cite{SafiJoyez} The nonlinearity appears even in absence of interactions due to coupling to the harmonic oscillator, and becomes  more pronounced in presence of interactions in the conductor.
The absorption noise becomes more structured at the frequency of the oscillator. While the excess FF noise is always positive in the SBS regime, this is not always the case in the WBS: less noise is absorbed by the TLL at the frequency of the oscillator when the system is undergoing a bias voltage.  
In this article we have also stressed the difference between the SBS and WBS regimes, where the back action is stronger.  We have shown that in the weak feedback regime, the role of the TLL liquid is essentially to damp in a classical way the harmonic oscillator. When the relaxation time and the period of the harmonic oscillator are on the same scale, interesting features of non-Markovian transport emerge and such system could be of interest to study its effects.

We acknowledge fruitful discussions with D. Estève, J. Gabelli, P. Joyez, O. Parlavecchio, F. Pierre, F. Portier, P. Roche, E. Sukhorukov. We would also like to thank M. Albert and P. Joyez for a careful reading of the manuscript.

\appendix

\section{Keldysh Green Functions}\label{app:green}
\begin{figure}[h!]
	\centering
	\begin{pspicture}(0,0)(7,1.3)
	\psset{gridcolor=green,
subgridcolor=yellow}
\rput[bl](.3,.31){\includegraphics[width=0.37\textwidth]{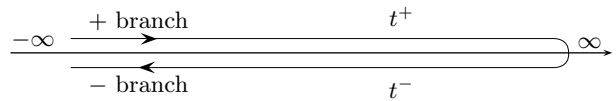}}
\rput[b](-.2,.61){$-\infty$}
\rput[b](7.2,.61){$\infty$}
\rput[b](1.2,.85){$+$ branch}
\rput[b](1.2,0){$-$ branch}
\rput[b](-.5,.55){\psline[linewidth=.5pt]{->}(8,0)}
\rput[b](1.2,.355){\psline[linewidth=1.5pt]{<-}(0.1,0)}
\rput[b](1.35,0.737){\psline[linewidth=1.5pt]{->}(0.1,0)}
\rput[b](4.7,.9){$t^+$}
\rput[b](4.7,-0.05){$t^-$}
\end{pspicture}
\caption{\label{keld}Keldysh Contour.}
\end{figure}

In the following, we compute Keldysh Green's functions for both a pure TLL and the LC resonator. They are labelled by two subscripts $\eta,\eta'=\pm$ which depend on whether each of the two time arguments is affected to the upper (+) or the lower branch (-) of the Keldysh contour (See Fig. \ref{keld}).

\subsection{Electronic Green function}\label{app:gel}
The hamiltonian of a Luttinger liquid is given by:
\begin{equation}
H_F=\frac{1}{2\pi}\int dx \;uK\left(\nabla\theta(x)\right)^2+\frac{u}{K}\left(\nabla\phi(x)\right)^2.
\end{equation}
From which we derive the imaginary-time Green function:
\begin{eqnarray}
&\CP_{\el}(\qq)=\cor{\theta^*(\qq)\theta(\qq)}=\frac{u\pi K^{-1}}{\omega_n^2+u^2k^2}, \\
&\CP_{\el}(x,\tau)=\frac{1}{4K}\ln\left(\frac{\exp(\tfrac{2\pi |x|}{\beta u})}{2\left[\cosh\left(\tfrac{2\pi x}{\beta u}\right)-\cos\left(\frac{2\pi\tau}{\beta}\right)\right]}\right).
\end{eqnarray}
By changing $K$ into $1/K$, we get the correlations of the conjugate field $\phi$. In order to obtain the real time Green functions, we have to perform the analytical continuation $\tau\to\ii t$. Note that for $|t|>|x|/u$, the term inside the log function is negative. Using $\ln(-x)=\ln|x|+\ii\pi$, we obtain the real time Keldysh Green functions $\CP_{\el}^{^{\eta\eta'}}(x,t)=\cor{T_K\theta^\eta(x,t)\theta^{\eta'}(0,0)}$ with $\eta,\eta'=\pm$ ($T_K$ is the Keldysh ordering operator) :
\begin{equation}
\begin{array}{rl}
\CP_{\el}^{^{\pm\pm}}(x,t)=&-\frac{1}{4K}\ln\left(\frac{\exp(\tfrac{2\pi |x|}{\beta u})}{2\left|\sinh^2(\tfrac{\pi t}{\beta})-\sinh^2(\tfrac{\pi x}{\beta u})\right|}\right)\\
&\mp\frac{\ii\pi}{4K}\Theta\left(|t|-\dfrac{|x|}{u}\right),\\
\end{array}
\label{eq:cpp}
\end{equation}

\begin{equation}
\begin{array}{rl}
\CP_{\el}^{^{\pm\mp}}(x,t)	=&-\frac{1}{4K}\ln\left(\frac{\exp(\tfrac{2\pi |x|}{\beta u})}{2\left|\sinh^2(\tfrac{\pi t}{\beta})-\sinh^2(\tfrac{\pi x}{\beta u})\right|}\right)\\
&\pm\frac{\ii\pi}{4K}\sgn(t)\Theta\left(|t|-\dfrac{|x|}{u}\right),
\end{array}
\label{eq:cpm}
\end{equation}
where $\Theta(t)$ is a Heaviside function. At zero temperature, we recover a $-\ln|(ut^2)-x^2|$ form. The imaginary part of these expressions is non zero when $t>|x|/u$. It translates the propagation of a charge density wave through the wire that emerges from a perturbation at $(x,t)=(0,0)$. This allows us to recover the retarded Green function: 
\begin{equation}
C_{\el}^R=C_{\el}^{++}-C_{\el}^{-+}=-\frac{\ii\pi}{2K}\theta\left(t-\frac{|x|}{u}\right),
\end{equation}
which is identical to the one of a sound wave propagating in a fluid. This is consistent with the d'Alembertian action given for the bosonic field. It is important to note that, since $\ln(-1)=\pm\ii\pi$, there is an indetermination of the sign of the complex term. This point is discussed at Eq.\eqref{eq:symmbreak}. 

\subsection{Oscillator Green function}\label{app:gosc}
We also give the Keldysh Green functions of a harmonic oscillator taken in its ground state defined by $\CP^{\eta\eta'}_{\res}(t)=\cor{T_K\varphi^\eta(t)\varphi^{\eta'}(0)}$ the following expressions:
\begin{equation}
\CP^{\pm\pm}_{\res}(t)=\rho\left(\coth(\frac{\beta\upsilon}{2})\cos(\upsilon t)\pm\ii\sin(\upsilon|t|)\right),
\end{equation} 

\begin{equation}
\CP^{\pm\mp}_{\res}(t)=\rho\left(\coth(\frac{\beta\upsilon}{2})\cos(\upsilon t)\pm\ii\sin(\upsilon t)\right),
\label{eq:GreenRes}
\end{equation} 
\subsection{Damped harmonic oscillator Green function}
The Green function of the damped harmonic oscillator in Matsubara frequencies is given by:
\begin{equation}
\CP_{\bos}(\omega_n)=\rho\dfrac{1}{\omega_n^2+2|\omega_n|/\tau_R+\Omega^2}
\label{eq:OscM}
\end{equation}
The real-time Green function, for $\tau_R\Omega>1$ is given by:
\begin{equation}
\CP_{\bos}^{\eta\eta}(t)=\frac{\rho\Omega}{\tilde\Omega}\left(e^{-|t|(\tfrac{1}{\tau_R}+\ii \tilde\Omega)}+\frac{2}{\pi} \Im g(|t|(\tilde\Omega+\frac{\ii}{\tau_R})\right)
\label{eq:ppg}
\end{equation} 
with $\tilde \Omega=\sqrt{|\Omega^2-\tau_R^{-2}|}$ and 
\begin{eqnarray}
&g(z)&=-C_i(z)\cos(z)-S_i(z)\sin(z),\\
&C_i(z)&=\displaystyle \int_z^{\infty}\frac{\cos(t)}{t}dt,\\
&S_i(z)&=\displaystyle \int_z^{\infty}\frac{\sin(t)}{t}dt.
\end{eqnarray}
The other two Green functions $\CP^{\eta-\eta}(t)$ are obtained by changing $\exp(\ii\tilde\Omega\eta |t|)$ to $\exp(\ii\tilde\nu \eta t)$ We see that owing to the absolute value of Eq.\ref{eq:OscM}, the real-time Green function has, in addition to the expected exponential decay, a term that embodies the feed-back effects of the Luttinger liquid on the brownian motion the harmonic oscillator. For $\rho\ll1$ this correction is negligible.
\section{Calculation of the  current and noise }\label{app:back}
	\subsection{The WBS limit}
	\subsubsection{Partition function}
We give here a brief summary of how the calculation of the backscattering current was performed. 
We define the generating function as:\cite{tztl}
\begin{eqnarray*}
&\MZ[J]&=\frac{1}{N_\MZ}\displaystyle\int \mathcal{D}\Phi^{\pm}\exp\left(-\frac{1}{2}\int d\rr \Bphi(\rr)^T\tilde{\MBC}^{-1} \Bphi(\rr)\right)\\
&&\displaystyle\times\exp\left(\sum_{\eta=\pm}-\ii\eta\iinfty dt\mathcal{H}_B[\phi^\eta]\right)\\
&&\displaystyle\times\exp\left(-\ii\int dt\boldsymbol{J}^TQ  \partial t \Bphi\right),
\end{eqnarray*}
where:
\begin{equation*}
\tilde{\MBC}= \begin{pmatrix} \CP^{++}(\rr,\rr') &\CP^{+-}(\rr,\rr')\\ \CP^{-+}(\rr,\rr') &\CP^{--}(\rr,\rr')\end{pmatrix}, \hspace{.5cm}
 Q= \begin{pmatrix} 1 &-1\\ 1 & 1\end{pmatrix},
\end{equation*}
and $\BJ=(eE(t)/\pi,J(t))$ with $E(t)$ the external electric field. At this stage it is possible to integrate out the modes at $x\neq0$, this is equivalent to setting $r=0$ in the Green functions.\cite{kf} Making the change of variable:
\begin{equation}
\tilde \Bphi=\Bphi-\ii\iinfty dt'\tilde\MBC(t,t')Q \partial_{t'}\BJ(t')
\end{equation}
we obtain:
\begin{eqnarray*}
&\MZ[J]&=\exp\left(-\frac{1}{2}\iint dtdt' J(t)^T\CP^K(t,t')J(t')\right)\\
&&\times\exp\left(\frac{e}{\pi}\iint dtdt'J(t)C^R(t,t')E(t')\right)\\
&&\times\cor{\displaystyle\exp\left(\sum_{\eta=\pm}-\ii\eta\iinfty dt'\mathcal{H}_B[\phi^\eta]\right)}
\end{eqnarray*}
		\subsubsection{Average current}
The backscattering average current is given by:
\begin{equation}
j_B(t)=\lim_{J=0}\frac{\updelta}{\updelta J(t')}\cor{T_K e^{\sum_{\eta=\pm}-\ii\eta\int dt'\mathcal{H}_B[\tilde\phi^\eta]}}.
\end{equation}
 One obtains, at second order in $v_B$:
\begin{eqnarray}
&\cor{j_B}=&2ev_B\left<T_K \sin(2\phi^+(0))+\sin(2\phi^-(0))\right.\nonumber\\
&&\left. e^{\sum_{\eta=\pm}-\ii\eta\int dt'\mathcal{H}_B[\tilde\phi^\eta]}\right>\nonumber\\
&						 & \approx 2\ii ev_B^2\iinfty dt  \langle T_K\left(\sin(2\phi^+_0)+\sin(2\phi^-_0)\right) \\
&		&\times\left(\left(\cos(2\phi^+_t+e^*Vt)-\cos(2\phi^-_t-e^*Vt)\right)\right)\rangle_0 \nonumber\\
&	  =& 2\ii e^*v_B^2\iinfty dt \sum_{\eta\eta'}\eta\sin(e^*Vt)e^{2\GP^{\eta\eta'}(t)},
\end{eqnarray}
where
\begin{equation}\label{eq:defG}
e^{2\GP^{\eta\eta'}(t)}=\cor{T_K e^{-2\ii\phi^{\eta}(t)}e^{-2\ii\phi^{\eta'}(0)}},
\end{equation}
and $\GP$ denotes the charge field Green function.
For $\eta=\eta'$, the Green functions are even with respect to time. For $\eta'=-\eta$, one finds:
\begin{equation}
e^{2\GP^{\eta-\eta}(t)}=\left(\tfrac{\pi\alpha}{\beta u}\right)^{2K}\frac{e^{\eta\ii K \pi\sgn(t)}}{\sinh^{2K}\left(\frac{\pi|t|}{\beta}\right)}
+\sum_{k=0}^{\lfloor K^{-1}\rfloor }C_k\delta^{(k)}(t)
\label{eq:}
\end{equation}
Its Fourier Transform gives the backscattering probability:
\begin{equation}
\iinfty dt e^{\ii\omega t}e^{2\GP^{-\eta\eta}(t)}=\Gamma_{\tfrac{1}{K}}(\eta\omega)+\sum_{k=0}^{\lfloor K^{-1}\rfloor }C_k \omega^k
\end{equation}
The last term arises from the kernel of $\sinh^{-2/K}(\pi t/\beta)$ and the $C_k$ are constants that have to be set. To understand the origin of these terms, it is useful to recall the Fourier transform of a Fermi distribution of electron or holes which read:
\begin{equation}
\frac{\beta}{2\pi}\iinfty dE f(\pm E)e^{\ii E|t|}=\pm\frac{\beta}{2}\delta(t)-\frac{\ii}{4\sinh\left(\frac{\pi |t|}{\beta}\right)}.
\label{eq:symmbreak}
\end{equation}  
By fixing the $C_k$, we are in fact breaking the particle-hole symmetry. Since the $\Gamma_K$ functions are proportional to the density of states of electrons, this is done for integers value of $K^{-1}$ by fixing the $C_k$ constants so that the following boundary conditions are fulfilled: $\Gamma(\infty)=0$ for electrons and $\Gamma(-\infty)=0$ for holes. For non integers value $K^{-1}$, this symmetry is already broken when the sign of the imaginary part in Eq.\eqref{eq:cpm} is chosen, therefore the previous boundary conditions are automatically fulfilled. Another non-trivial check is the general equation $\CP^K(\omega)=\coth(\beta\omega/2)(\CP^R(\omega)-\CP^A(\omega))$ that also fixes the sign of the analytical continuation and ensures that we are dealing with electrons and not holes.

		\subsubsection{FF Noise}
To compute the average symmetrized current one can derive the partition function twice with respect to the field $J$.\cite{tztl} To obtain the non symmetrized noise, one has to perform the same calculation but by introducing $\BJ=(J^+,J^-)$ and then derive the partition function with respect to these two fields to recover equation (\ref{eq:sagen}) where we are left to compute the average value of $\cor{j^+j^-}$ which yields, for a TTL:
\begin{eqnarray}
&\cor{T_K j^+j^-}(\omega)&=(ev_B)^2\iinfty dt e^{\ii\omega t} \cos(eVt)e^{2\GP^{+-}(t)}\\
&&=(ev_B)^2\sum_{\eta=\pm}\Gamma_{\tfrac{1}{K}}(-\omega+\eta eV)
\end{eqnarray}
Using the balance equations $\Gamma(-eV)=e^{-\beta eV}\Gamma(eV)$, one finally obtains:
\begin{equation}
\cor{T_K j^+j^-}(\omega)=e\sum_{\eta=\pm}{\cal N}(\omega+\eta eV)I_B(\omega+\eta eV)
\label{eq:}
\end{equation}
	\subsection{The tunneling limit}

In the tunneling limit, the results are straightforward. By defining the current operator as in Eq.\ref{eq:jt}, one gets for a Luttinger liquid:
	\begin{eqnarray}
		&\cor{j_T}&=ev_T^2\iinfty dt \sin(eVt)\cor{T_K e^{2\ii\theta^+}e^{-2\ii\theta^-}}\\
		&\mathcal I_T&=ev_T^2\left(\Gamma_K(eV)-\Gamma_K(-eV)\right)
\end{eqnarray}
For the average noise, the result is much simpler that in the WBS limit as there is only one noise source. The non-symmetrized shot noise is obtained by computing  $\cor{T_K j^+j^-}$, which yields:
\begin{eqnarray}
&\cor{T_K j^+j^-}(\omega)&=(ev_t)^2 \displaystyle\int dt e^{\ii\omega t}\cos(eVt)\cor{T_K e^{2\ii\theta^+}e^{-2\ii\theta^-}} \nonumber\\
&&=e\sum_{\eta=\pm}{\cal N}(\omega+\eta eV)I_T(\omega+\eta eV).
\end{eqnarray}

\bibliographystyle{apsrev4-1}
\bibliography{Biblio}
\end{document}